\documentclass[aps,prd, preprint,nofootinbib]{revtex4-1}
  \usepackage{graphicx,psfrag,epsfig}
\newcommand{\beq}{\begin{equation}}
\newcommand{\eeq}{\end{equation}}

\newcommand{\bea}{\begin{eqnarray}}
\newcommand{\eea}{\end{eqnarray}}

\begin{document}
\begin{flushright}
UMD-PP-13-001\\
May 2013\
\end{flushright}
\vspace{0.3in}

%\documentclass{appolb}

% epsfig package included for placing EPS figures in the text
%------------------------------------------------------

%%%%%%%%%%%%%%%%%%%%%%%%%%%%%%%%%%%%%%%%%%%%%%%%%%
%                                                %
%    BEGINNING OF TEXT                           %
%                                                %
%%%%%%%%%%%%%%%%%%%%%%%%%%%%%%%%%%%%%%%%%%%%%%%%%%
%\begin{document}
% \eqsec  % uncomment this line to get equations numbered by (sec.num)
\title{Possible Implications of Asymmetric Fermionic Dark Matter for Neutron Stars  %
  %\thanks{Presented by I.~Goldman at the XXXV International Conference of Theoretical
%Physics, "Matter to the Deepest", Ustron, Poland, September 12-18, 2011.
}
\author{I. Goldman$^a$, R. N. Mohapatra$^b$, S. Nussinov$^{c,d}$, D. Rosenbaum$^e$  and V.~Teplitz$^{e,f}$}
\affiliation{ $^a$Department of Exact Sciences,  
Afeka Tel Aviv Academic Engineering College, Tel Aviv, Israel }
\affiliation{$^b$Maryland Center for Fundamental Physics and Department of Physics, University of Maryland, College Park, USA} 
\affiliation{$^c$School of Physics and Astronomy, Tel Aviv University, Tel Aviv, Israel } 
\affiliation{$^d$Schmid College of Science,Chapman University, Orange, California 92866, USA}
\affiliation{$^e$ Physics Department, Southern Methodist University, Dallas}
\affiliation{$^f$NASA Goddard Space Flight Center, Greenbelt, MD, USA} 
%}

\begin{abstract}
  We consider the implications of  fermionic asymmetric dark matter for a ``mixed neutron star''  composed of ordinary baryons  and dark fermions.  We find examples, where for a certain range of dark fermion mass -- when it is less than  that of ordinary baryons -- such systems can reach  higher masses than the maximal values allowed for ordinary (``pure") neutron stars. This is shown both within a simplified, heuristic Newtonian analytic framework with non-interacting particles and via a  general relativistic numerical calculation, under certain assumptions for the dark matter equation of state. Our work applies to various dark fermion models such as mirror matter models and to other models where the dark fermions have self interactions.

\end{abstract}
%\PACS{95.35.+d, 97.60.Jd, 26.60.Kp, 04.20.-q}
 \maketitle  
\section{Introduction}

Cold dark matter (CDM),   favored by most astrophysical and cosmological  observations, can be realized in symmetric or  asymmetric scenarios.
In the first class of models, dark matter is made of stable $X$  particles and an equal amount of stable $\bar X$ antiparticles of mass $m_X$. In the early universe, these were   in thermal equilibrium and their residual abundance $\Omega_X$ is fixed, at the ``freeze-out" value, when the rate of the Hubble expansion overcomes that of $\bar X-X $ annihilation. A prototypical example, which has been extensively studied, is provided by supersymmetric models with R-parity conservation, where the lightest superpartner is stable and plays the role of the dark matter of the universe.  
 
 As yet no  light sub-TeV SUSY partners  have been discovered at the LHC,
and searches for electrons, positrons or photons from annihilations  in clumps of DM in and around our Galaxy,  do not provide solid  ``indirect" evidence for symmetric massive  DM. Moreover, the ongoing direct underground searches put very strong bounds on the scattering cross sections of massive X's  on nuclei. In  the symmetric case,  accretion of  DM particles onto the sun accelerates  the rate of particle-antiparticle annihilation. The resulting photons, electrons, etc., are all trapped in the star. However, for massive DM particles, looking  in ICE-CUBE (the large km$^3$ Cerenkov radiation detector near the south-pole) for the resulting UHE neutrinos is an excellent indirect detection method. The fact that no such energetic neutrinos have been detected constrains symmetric DM models.
 Consequently there has been, in recent years, a renewed interest  in a second class of models: the asymmetric dark matter  (ADM) models.
In such models,  the relic  ADM density is determined in a manner  analogous to that of ordinary baryonic matter, not by the freeze-out of DM annihilation, as in the symmetric case. An excess of   dark fermions (over the antifermions) remains after the annihilation of most antiparticles. The required dark matter density in such models is readily achieved if the ratio of the  ADM particles mass and that of ordinary baryons is tuned inversely with the corresponding ratio of asymmetries. Many examples  of such  models have been proposed over the years \cite{nuss85}. 
Here we consider variants in which the dark matter particle is rather light with mass in the sub- GeV range. Scattering of such light CDM on most detector materials yields recoil energies $\sim 0.1$ KeV which are below the existing experimental thresholds. Hence the present upper bounds on the X-N scattering cross-sections do not apply. Also the stringent indirect upper bounds from missing energy searches at the collider \cite{Harnik} apply for massive mediators of the X-nucleon interactions - and do not apply if the exchange of a  relatively light  "dark photon"  mediates X-N scattering, as is the case in several asymmetric dark matter models.   This may allow $\sigma_{X-N}$   of order $10^{-34}\ cm^2$  -- which is high enough to be relevant in astrophysical settings and yet is 10 orders of magnitude smaller than the intra- species cross-sections of ordinary matter and potentially of dark matter.

For our purpose, in this paper, it is useful to consider a class of models for ADM proposed in \cite{an} and its possible variants. These contain an additional sector mirroring  our universe.  The mirror sector  consists of particles and forces 
  related to those of the familiar standard model by a mirror symmetry. As a result, there are no new  parameters in the model prior to gauge symmetry breaking \cite{okunreview}. In generic mirror models, an important constraint comes from BBN due to the presence of three extra neutrinos and an extra photon. One way to avoid this constraint is to to assume that the temperature of the mirror sector is smaller than that of the familiar sector \cite{zurab}. An alternative possibility detailed in  \cite{an} is to   have the symmetry breaking in the mirror sector sufficiently different from that in  the familiar sector so that all the mirror neutrinos and mirror photon are heavy and  have decayed by the BBN epoch and only the mirror neutrons survive constituting the dark matter.  Our considerations are independent of the model details of \cite{an} and could be applied to variants of the model where the mirror photon is very light,  e.g. less than an eV. 
The details of large scale structure formation depend on the specific model for the asymmetric dark matter. Being self interacting  the dark matter  will  no longer provide collision-less dark halos with many possible cosmological ramifications~\cite{haibo}.

In  asymmetric DM models,  the dark matter particles   can accumulate in astrophysical objects and alter their properties.  The goal of the present paper is to study the effect of such accumulation on neutron star properties. Similar studies for the case of scalar ADM have been reported in several papers \cite{yu}, where Bose condensation plays an important role. The situation for the case of fermionic dark matter is however very different due to the Pauli exclusion principle and our goal is to make some remarks on this case. We find that under certain conditions, the mass of the mixed neutron- DM star can exceed the Chandrasekhar-like mass limit for ordinary neutron stars. 
 The recent discovery of a  $2 M_{\odot}$  binary radio pulsar \cite{2mo},  already severely constrains nuclear matter equations of state, if it is a ``pure" neutron star and possible future observation of  such neutron stars  with  higher masses  would be very difficult to reconcile with standard hadronic physics  but, as we show in this paper, such higher mass neutron stars seem to be more easily realized as  mixed neutron stars.
 
 Another  result of our discussion is that, for a mixed neutron star with two species which interact with each other only via gravitational interactions, requiring stability (see sec. 4) imposes an interesting scaling relation between the number and energy density and pressure. Such a relation will constrain the density profiles of the model as well as the number distribution of the two species.
 
 This paper is organized as follows: in sec. II, we discuss a Newtonian model for a mixed neutron star; in sec. III, we present the general relativistic treatment of the mixed neutron star  containing both ordinary and dark fermions.  In sec. IV, we discuss the implications of stability (extremum with respect to variations of mass-energy density keeping the total number of particles fixed) for the mixed neutron star; see Eq. (38). We find the interesting relation cited above among pressure, density and the particle number density in the two sectors. In sec. V, we present an illustrative example where we employ the same equation of state for the familiar sector and the dark sector to discuss the impact on neutron star mass. In sec. VI, we present some astrophysical discussion and we conclude in sec. VII.

\section{Maximal Mass of mixed neutron stars:  heuristic discussion} 
 
Before proceeding to a detailed   analysis, let us start with a heuristic discussion based on Newtonian intuition for a mixed neutron star ignoring nuclear physics effects. We generalize to the present mixed case, the discussion in \cite{Lightman}   which estimated the maximal mass of an ordinary neutron star. 
The total energy of a mixed neutron star is the sum of the relativistic  Fermi energy and the gravitational energy: 
$E= E_F + E_G$ where:  
\begin{equation}
E_F=   \beta \frac{\hbar c}{R_1}N_1^{4/3} +  \beta \frac{\hbar c}{R_2}N_2^{4/3} \ , \ \ \ \ \ \  E_G= - \frac{1}{8 \pi  G}\int_0^{\infty}  \left( \frac{G m(r)}{r^2}\right)^2 4\pi  r^2 dr
\end{equation}

Here, $\beta$ is a coefficient of order unity, $N_{i}$ ($i=1,2$) denote the total number of baryons ($i=1$) and dark fermions ($i=2$) and $m(r)$ is the mass enclosed within a sphere of radius $r$. We use a Newtonian approximation for $ m(r)$ and assume further ( as justified a posteriori by the detailed numerical calculations)
that the energy and mass densities  of the two species can be approximated as constants, in their respective spheres of radii $R_1,\ R_2$ .
 Thus, we have
\begin{eqnarray}
m(r)=M_1 \left( \frac{r }{R_1}\right)^3   + M_2 \left( \frac{r }{R_2}\right)^3 , \ \ \  \ r \leq R_1\\
m(r)= M_1 +  M_2 \left( \frac{r }{R_2}\right)^3, \ \ \  \ R_1 \leq r \leq R_2\\
m(r)=  M_1 + M_2 \ , \ \  \ r \geq R_2
\end{eqnarray} 

Substituting the above expression for $m(r)$ in the integral expressing  $E_G$ and performing the integration over the inner, the intermediate and the outer regions we find:
\begin{equation}
E_G = -\frac{3}{5} \frac{G M_1^2}{R_1}  -\frac{3}{5} \frac{G M_2^2}{R_2}  - \frac{3}{2}  \frac{G M_1 M_2}{R_2}+ \frac{3}{10} \frac{G M_1 M_2}{R_2}\left(\frac{R_1}{R_2}\right)^2
 \end{equation}

For the case where  $R_2$ exceeds $R_1$, the last term is small compared to the  previous term and can be neglected.  Hence, we get 
 \begin{equation}
 E=  \beta m_{1}^{-4/3} \frac{\hbar c}{R_1}M_1^{4/3}- \frac{3}{5} \frac{G M_1^2}{R_1}+ \beta m_{2}^{-4/3} \frac{\hbar c}{R_2}M_2^{4/3}- \frac{3}{5} \frac{G( M_2^2 + 2. 5 M_1 M_2)}{R_2}
 \end{equation}
 where in the spirit of the Newtonian approximation we used 
$ M_i = N_i m_{i}, \ \ i=1,\ 2$, where $m_{1,2}$ are the masses of the familiar neutron and the dark fermion respectively.
 
Following \cite{Lightman}, one can argue that the sums of the coefficients  multiplying $1/R_1$ and $1/R_2$ should be positive in order to avoid gravitational collapse to a black hole.
Thus one gets
\begin{equation}
\label{analytic_constr11}
M_1 - M_0 \leq  0  
\end{equation}
and
\begin{equation}
\label{analytic_constr12}
M_2 ^2 - M_2^{4/3} M_0^{2/3}\left(\frac{m_{1}}{m_{2}} \right)^{4/3} + 2.5 M_1 M_2 \leq 0 
\end{equation}

with
\begin{equation}
\label{ M0}
 M_0 = \left(\frac{5 \beta}{3}\right)^{3/2}\frac{m_{pl}^3}{m_{1}^2}
\end{equation}
denoting the maximal mass of a pure neutron star. Its value  is of the order of a solar mass. Some realistic  nuclear equations of state for familiar neutrons, including the one employed in the next sections,  
allow values of $M_0\approx 2.5 M_{\odot}$. 

  The total mass of the mixed neutron star is 
\begin{equation}
M= M_1 + M_2
\end{equation}

The above constraints correspond to the case where the neutrons-sphere is within the outer radius, $R_2$. We can consider the opposite case in which the dark matter sphere is enclosed within the neutron-sphere. In this case we will get
\begin{equation}
\label{analytic_constr21}
M_2 - M_0\left(\frac{m_{1}}{m_{2}} \right)^2 \leq  0 
\end{equation}

\begin{equation}
\label{analytic_constr22}
M_1 ^2 - M_1^{4/3} M_0^{2/3}  + 2.5 M_1 M_2 \leq 0 
\end{equation}
  where $ M_0$    
   is the same as defined above.
   
\begin{figure}[h]
\centerline{\includegraphics[scale=0.8]{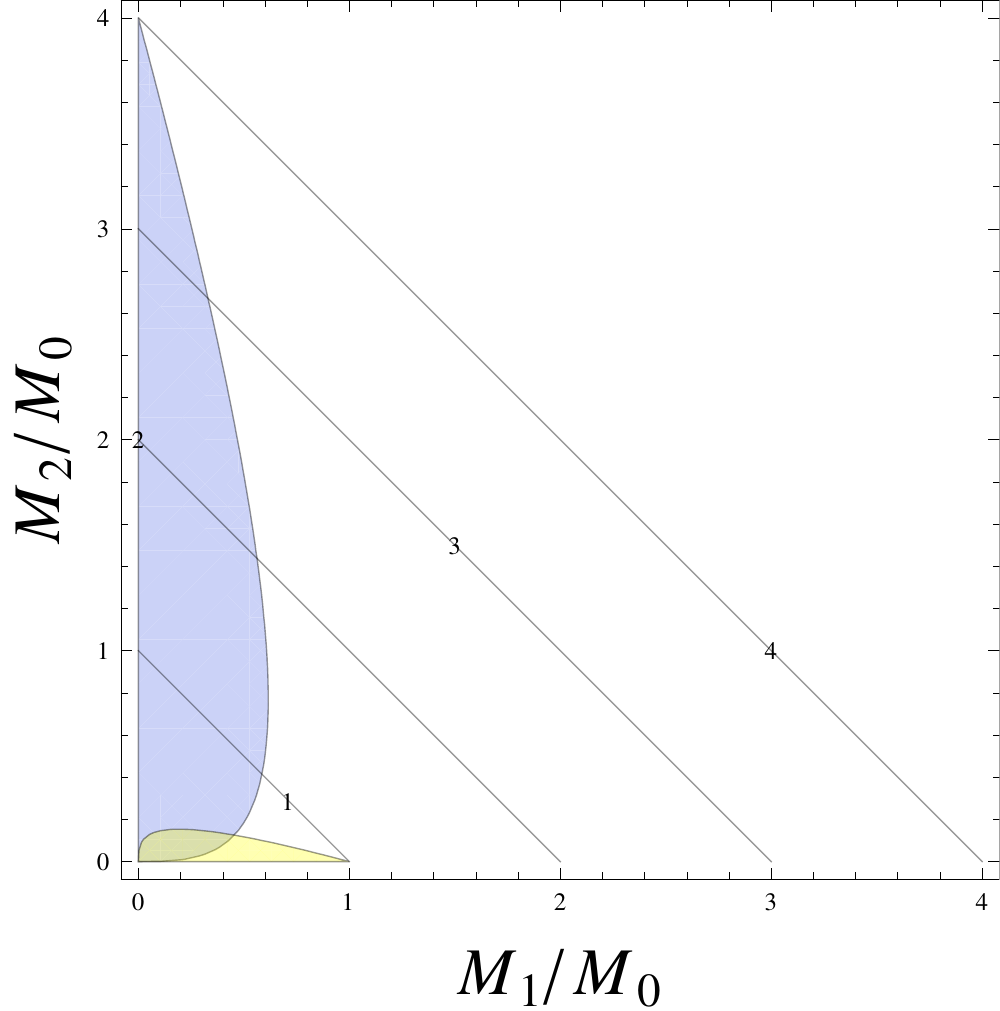}}
\caption{Allowed  regions in the    $M_1-M_2$ plane for $m_{2}= \frac{1}{2} m_{1}$. {\it Blue shaded area}:  $R_2> R_1$. {\it Yellow shaded area}:   $R_1> R_2$. { \it Diagonal lines:} loci of total mass.} 
\label{analytic}
\end{figure}

Note that we do not obtain constraints on the radii, because the relativistic limit for the Fermi energies was adopted. Had we taken the general expression for the Fermi energies and minimized with respect to each radius, we would have obtained also constraints on the radii.
However, this would have increased the complexity of the  heuristic analytic estimates. Furthermore we find a similar result when solving numerically the general relativistic (GR) 
mixed star model.
It follows from the above constraints   that  the maximal mass of an ordinary neutron star is $M_0$ and that of a pure dark (mirror) analog of a neutron  star is  $ M_0 (m_{1}/m_{2})^2$. Therefore, this mass will be larger than that of an familiar neutron star only if the mass of the dark (mirror) baryon is  smaller than the mass of the neutron. Since having degenerate neutron and dark matter fermion mass decreases the maximal mass of the neutron star, this holds {\it a~forteriori}  for  mixed neutron stars.
In what follows we use  an illustrative value of $m_{2}= \frac{1}{2} m_{1}$.
The blue shaded area in Fig. \ref{analytic} marks the region in the $M_1-M_2$ plane allowed by   equations (\ref{analytic_constr11}, \ref{analytic_constr12}) for $m_{2}= \frac{1}{2} m_{1}$. The yellow shaded area  marks the region in the $M_1-M_2$ plane allowed by   equations (\ref{analytic_constr21}, \ref{analytic_constr22}).  The masses are expressed in units of   $M_0$. 

The blue shaded area corresponds to the case where the dark matter sphere extends beyond that of the familiar neutron sphere.
%either the case where a dark neutron star accreted ordinary matter or to the previous case when the accreted dark mass is too large to be confined inside the neutron sphere and the configuration switches to that of the neutron sphere confined within the dark matter sphere. 
This region  is particularly interesting, as it allows a total mass exceeding the maximal mass of a pure neutron star.
Note  that  the maximally allowed dark fermion mass is always smaller than  the neutron mass for total mass to exceed the  Chandrasekhar limit, $M_0$ with the ``ultimate'' limit on the mixed neutron star mass being four times that of the pure neutron star for dark matter mass being half that of the familiar neutron. Clearly the ultimate limit depends on the ratio $m_2/m_1$.
%The  exception   seen  at the low end of the dark masses in this region is
%probably due to our approximation that in this corner allows artificially $R_2> R_1$ while
%$M_2<M_1$.
% \newpage

\section{Dynamics of a mixed neutron star}
%\subsection{The model euqations}
Encouraged  by the above heuristic results, we proceed to a fully general relativistic (GR) discussion. The energy momentum tensor of a mixture  of  two non-interacting ideal fluids:
 \begin{equation}
\label{tmunu}
T^{\mu  \nu} = T_1^{\mu  \nu}+ T_2^{\mu  \nu} =( \rho_1 +p_1) u_1^{\mu}u_1^{\nu} - p_1 g^{\mu \nu} +( \rho_2 +p_2) u_2^{\mu}u_2^{\nu} - p_2 g^{\mu \nu}   
\end{equation} 
%As argued in  the introduction, any   inter-species interaction must be weaker than the intra-specie interaction by a factor $\geq 10^{10}$.  
 We consider here an ideal case in which the interaction between DM and familiar fermions is sufficiently weak  that each tensor  $T_{1,2}$ is separately  conserved.

We   look for a spherically symmetric static solution of the Einstein field equations  for the two-fluid ``mixed neutron star".   Since we address here a mixture of two fluids, we rederive the hydrostatic equilibrium equations for this case,  starting with the Einstein field equations\cite{Goldman_11}. 
 The line element squared of a spherically symmetric static metric can be written in the Schwarzshild coordinates $(t, r, \theta,\phi)$ as
 
\begin{equation}
\label{metric}
 ds^2=g_{\alpha \beta}dx^{\alpha} dx^{\beta}=  e^{ 2\phi(r)}c^2 dt^2 - e^{ 2\lambda(r)}dr^2 - r^2 \left(d \theta^2 + \sin^2(\theta) d\phi^2\right)
\end{equation}    
The 
  $\alpha=\beta=t$ and the  $\alpha=\beta=r$ equations, respectively, are   
 
\begin{equation}
\label{tt} 
\frac{1}{r^2}\frac{d}{dr}\left(r(1- e^{-2\lambda(r)})\right) = 8 \pi \frac{G }{c^{2}}T^t_t = 8 \pi \left( \rho_1 +  \rho_2  \right) 
\end{equation} 

    \begin{equation}
 \label{rr}
-r^{-2}+ e^{-2\lambda(r)}\left(r^{-2}+ 2 r^{-1} \frac{d\phi(r)}{dr}\right)  = 8 \pi \frac{G }{c^{2}}T^r_r = - 8 \pi \frac{G }{c^{2}}( p_1 +  p_2)
\end{equation} 

We also have the separate two covariant conservation equations of  the energy momentum tensors:

 \begin{equation}
\label{conservation}
{T_i^{ \mu  \nu}}_{;_\nu }=0, \ \ \ i =1, 2
\end{equation} 

 The field equation (\ref{tt}) immediately yields 
 
 \begin{equation}
\label{lambda } 
e^{- 2\lambda(r)}= \left( 1- 2\frac{G }{c^{2}}\frac{m(r)}{r}\right) 
\end{equation} 

where  $m(r)$ is the mass enclosed within   $r$ and is  given by
 \begin{equation}
\label{mass} 
   m(r) = \int_0^r 4\pi \bigg(( \rho_1(r')  + \rho_2(r') \bigg) r'^2   dr'
\end{equation} 

The two covariant conservation   equations of  the energy momentum tensors, equation (\ref{conservation}) take the form:

 \begin{equation}
\label{phi1}
   \frac{d\phi(r)}{dr}=  - \bigg(\rho_1(r) + p_1(r)\bigg)^{-1} \frac{d p_1(r)}{dr} 
\end{equation}
 \begin{equation}
 \label{phi2}
\frac{d\phi(r)}{dr} =  - \bigg(\rho_2(r) + p_2(r)\bigg)^{-1} \frac{d p_2(r)}{dr}
\end{equation}

Combining them with the field equation (\ref{rr}) leads to 
  a hydrostatic equilibrium equation for each of the species:

\begin{equation}
\label{equil1}
      \frac{d p_1(r)}{dr}= - G \bigg(\rho_1(r) + p_1(r)\bigg)\frac{m(r) + 4\pi r^3 \bigg(p_1(r)+ p_2(r)\bigg)}{r\bigg(r - 2 G c^{-2} m(r)\bigg)}
\end{equation} 

\begin{equation}
\label{equil2}
      \frac{d p_2(r)}{dr}= - G \bigg(\rho_2(r) + p_2(r)\bigg)\frac{m(r) + 4\pi r^3 \bigg(p_1(r)+ p_2(r)\bigg)}{r\bigg(r - 2 G c^{-2} m(r)\bigg)}
\end{equation} 

  These equations imply  that  each  fluid satisfies its own hydrostatic equilibrium equation  which is of the form of a  modified TOV  (Tolman, Oppenheimer, Volkoff) equation \cite{t}, \cite{ov}.  In this paper, we assume that the two fluids are  are coupled only through gravity. %the total mass $m(r)$ and the total pressure $p_1(r) + p_2(r)$.
 
 Given the two equations of state,  and 
  the two central energy densities,  the  
 TOV equations (\ref{equil1}, \ref{equil2}) are integrated up to $r=R_1$ where $p_1(R_1) = 0$. Species 1 is confined within this radius. From this radius on,  only TOV2 (equation \ref{equil2}) is integrated out to the radius $R_2$ where  $p_2(R_2) = 0$, which is the outer radius of the complete mixed neutron star.
 
 Once the solutions of equations (15-16) are obtained, equations (20-23) are solved with the boundary condition
 $\phi(R_2)=\frac{1}{2} \ln { \left( 1- 2\frac{G }{c^{2}}\frac{M}{R_2}\right)}$ with
 $M= m(R_2)$ being the mass of the mixed neutron star. This boundary condition is imposed by demanding that the inner solution be matched to the external   Schwarzschild solution for which  $\phi = -\lambda$. 
 
In this way, a two-parameter (namely the two central densities) family of static models is obtained. In contrast, ordinary neutron star models form a one-parameter (one central density) family of solutions.

It is interesting that,  combining  the two covariant conservation laws of fermion number densities 

\begin{equation}
\label{baryon_conservation}
\left( n_i u_i^{\alpha}\right) {;_\alpha }=0, \ \ \ i =1, 2
\end{equation}
with the two covariant conservation equations of  energy momentum tensors (\ref{conservation}) 
 we find the isentropic relations for each of the fluids

     \begin{equation}
\label{isentrop }  
 \frac{d \rho_i} { \rho_i  +p_i} = \frac{dn_i }{n_i},  \ \ \ \ \ i=1, 2
\end{equation}

Using Eq.(20, 21) \cite{Goldman 89}, and noting that $p_i(R_i) =0$,  these   equations lead to 
\begin{equation}
\label{constr}
 n _1(r)  =  m_{1}^{-1} e^{-\phi(R_1)+\phi(r)} \left(\rho_1(r) + p_1(r)\right) , \ \ \ \ \  n _2(r)  =  m_{2} e^{-\phi(R_2)+ \phi(r)} \left(\rho_2(r) + p_2(r)\right)
\end{equation}
In turn these constraints imply that inside the inner neutron radius $R_1$, where two species coexist and we have the relation,

\begin{equation}
e^{+\phi(R_1)}\frac{ n _1(r)m_{1}}{\rho_1(r) + p_1(r)}= e^{+\phi(R_2)}\frac{ n _2(r)m_{2}}{\rho_2(r) + p_2(r)}
\end{equation}
%Moreover, since $e^{\phi(R_1)- \phi(R_2)}<1$, specie 1, the heavier ordinary neutron, is less relativistic than specie 2, throughout the inner neutron radius $R_1$. 

  \section{ Stability of  the mixed neutron star}
 In what follows we show that the mass stability theorem \cite{Harrison} summarized in Weinberg's book \cite{weinberg}
can be extended for a mixed neutron star (and in effect more generally for any mixed star constructed out of two non-interacting fluids). This imposes constraints on the
equilibrium mass and density distributions of the dark matter relative to familiar neutrons in a neutron star.

\noindent{\underline {\bf  Theorem}} 
 
For quasistatic spherically symmetric structures with fixed given total baryon numbers (of the neutrons and the the dark mirror baryons), the total mass is stationary for variations of the the two energy densities $\rho_1(r),\ \rho_2(r)$ if  and only if the two equilibrium equations are satisfied. 

As in the case of a single fluid, one considers the case where the entropy per baryon is uniform.
Using the Lagrange multipliers method we explore the implications of demanding that
\begin{eqnarray}\label{A1}
\delta M - \lambda_1 \delta N_1 -  \lambda_2 \delta N_2 =0 %\eqno(A1)
\end{eqnarray}
where
$$M =\int_0^{\infty} 4 \pi r^2\bigg(\rho_1(r) +  \rho_2(r)\bigg) dr $$
and
$$N_ i=  \int_0^{\infty} 4 \pi r^2 n _i(r) \left[1- \frac{2 G m(r)}{c^2 r}\right]^{-1/2}  dr\ , \ \ \ \ i=1, 2  $$
with $m(r)$  the mass enclosed within the radius $r$ given by
$$m(r) = \int_0^ r 4 \pi r'^2\bigg(\rho_1(r') +  \rho_2(r')\bigg) dr'  $$

Combining the above equations, one gets
\begin{eqnarray}
 \int_0^{\infty} 4 \pi r^2\bigg(\delta \rho_1(r) +  \delta\rho_2(r)\bigg) dr -  \\ \nonumber \int_0^{\infty} 4 \pi r^2\bigg( \lambda_1  \delta 
 n_1(r) + \lambda_2 \delta
 n_2(r) \bigg) \left[1- \frac{2 G m(r)}{c^2 r}\right]^{-1/2}  dr \\ \nonumber
   -   G/c^2 \int_0^{\infty} 4 \pi r  \bigg( \lambda_1 n_1(r) + \lambda_2 n_2(r) \bigg)  \left[1- \frac{2 G m(r)}{c^2 r}\right]^{-3/2} \delta m(r) dr =0  \end{eqnarray} 
 
  The uniform entropy conditions, for each of the fluids,
    $ \frac{d \rho_i} { \rho_i  +p_i} = \frac{dn_i }{n_I},  \ \ \ \ \ i=1, 2 $ 
   imply that
\begin{eqnarray}\frac{\delta n_i(r)}{n_i(r)}= \frac{\delta\rho_i(r)  }{\rho_i(r) + p_i(r)} , \ \ \ \ i=1, 2 \end{eqnarray} 
 %The variation of $m(r)$ is

Substituting $\delta m(r) = \int_0^ r 4 \pi r'^2\bigg(\delta\rho_1(r') + \delta \rho_2(r')\bigg) dr'  $, we can rewrite Eq. (29)
 using the variations: $ \delta\rho_1(r), \delta\ \rho_2(r)$:
\begin{eqnarray}
 \int_0^{\infty} 4 \pi r^2 dr \left[\bigg(\delta \rho_1(r) +  \delta\rho_2(r)\bigg) dr - \bigg(\frac{ \lambda_1
 n_1(r) \delta\rho_1(r)  }{\rho_1(r) + p_1(r)} + \frac{\lambda_2  n_2(r) \delta\rho_2(r)  }{\rho_2(r) + p_2(r)} \bigg)B^{-1/2}  \right] \\\nonumber
  - G/c^2\int_0^{\infty}4 \pi r^2 \left\lbrace\int_r^{\infty} 4 \pi r'  \bigg( \lambda_1 n_1(r') + \lambda_2 n_2(r') \bigg) B^{-3/2}d r' \right\rbrace \bigg(\delta\rho_1(r) + \delta \rho_2(r)\bigg) dr     =0   \end{eqnarray} 
  where $B= \left[1- \frac{2 G m(r)}{c^2 r}\right]$.
In the last term, the integration order of $r$ and $r'$ was interchanged and the names where interchanged too. Since the variations  $ \delta\rho_1(r), \delta\ \rho_2(r) $ are arbitrary, equation (31) implies that in the region  $r \leq  R_1$, where the two species coexist

\begin{eqnarray} 1 -\lambda_1 a_1(r)  -\lambda_1 b_1(r)  - \lambda_2 b_2(r)=0 \ \ \ , \ \ \ \  1 -\lambda_2 a_2(r)  -\lambda_2 b_2(r)  - \lambda_1 b_1(r)=0 \end{eqnarray}
where

$$a_i(r)=\bigg( \frac{n_i(r)  }{\rho_i(r) + p_i(r)} \bigg) \left[1- \frac{2 G m(r)}{c^2 r}\right]^{-1/2}\ , \ \  \ \ i=1, 2 $$

$$b_i(r)=\frac{G}{c^2} \int_r^{\infty} 4 \pi r'    n_i(r')    \left[1- \frac{2 G m(r')}{c^2r'}\right]^{-3/2}d r'\ , \ \  \ \ i=1, 2 $$

Equation (32) implies that
\begin{eqnarray}
\lambda_1 a_1(r) = \lambda_2 a_2(r)  \end{eqnarray} 

so that

$$ 1 - \lambda_1\left(a_1(r) + b_1(r) +\frac{a_1(r)}{a_2(r)}b_2(r)\right) = 0 $$ 

which in turn implies,  using the facts that $\lambda_1$ and  $ \frac{a_1(r)}{a_2(r)}$are constants   

$$a_1(r)' + b_1 (r)' + \frac{a_1(r)}{a_2(r)}b_2 (r)'=0 $$ 

with  a prime denoting an $r$-derivative.

Using the above equations  results in

\begin{eqnarray}
p_1(r)' =  -\frac{ G}{c^2} \bigg(\rho_1(r) + p_1(r)\bigg)\frac{m(r) + 4\pi r^3 \bigg(p_1(r)+ p_2(r)\bigg)}{r^2 \bigg(1 - \frac{2 G  m(r)}{c^2 r}\bigg)}\end{eqnarray}
similarily we get
\begin{eqnarray}
p_2(r)' =  - \frac{ G}{c^2} \bigg(\rho_2(r) + p_2(r)\bigg)\frac{m(r) + 4\pi r^3 \bigg(p_1(r)+ p_2(r)\bigg)}{r^2 \bigg(1 - \frac{2 G  m(r)}{c^2 r}\bigg)}\end{eqnarray}
In the region  $r > R_1, \rho_1$ and its variation are zero. Therefore,
  instead of Eq (32) one gets

\begin{eqnarray}1-\lambda_2 a_2(r) - \lambda_2 b_2(r) =0\end{eqnarray}

implying

\begin{eqnarray}a_2(r)' + b_2(r)' =0\end{eqnarray}
which leads  again to Eq (35), the equilibrium structure equation for
species 2.

Equations (33) and (34) are the equilibrium structure equations that follow from the Einstein field equations.
Thus, these equations are the necessary and sufficient conditions for the mass of the star to be stationary under
arbitrary variations of the energy densities of the two species. In case that the second order variations are not zero,
the stationary point is only an extremum: either a minimum  implying a stable configuration  or a maximum implying an unstable equilibrium.

In addition, equation (33) implies that  in the region where the two
species coexist
\begin{eqnarray} \frac{n_1(r)  }{\rho_1(r) + p_1(r)}= Constant \ \frac{n_2(r)  }{\rho_2(r) + p_2(r)}\end{eqnarray}
which was obtained earlier (see Eq. (27)) as a result of the field equations and the covariant conservation laws. 

%It is also worth pointing out that if we start with a set of initial conditions that do not satisfy Eq.(35), gravitational interactions will bring them to equilibrium within a time scale given by
%$t\sim \frac{d^{3/2}}{\sqrt{G_N M}}$ where $d$ denotes the initial distance between the center of mass of the two clumps and $M$ , the mass of one of them. Putting typical neutron star mass and distance $\sim 10$ km, we find this time to be $\sim 10^{-4}$ sec. Thus, we believe that the above equation relating $n_i(r), \rho_i(r), p_i(r)$ needs to be satisfied as long as all other interactions are neglected and only gravity is playing a role.

    \section{An illustrative example}
 In discussion of neutron star equation of state(EOS), the role of nuclear forces is clearly important and is always an integral part of the discussion.
We present here an  example, where
we employ the nuclear equation of state of Steiner, Lattimer and Brown~\cite{steiner}.
   \begin{figure}[ht]
\begin{center} $
\begin{array}{cc}
\includegraphics[scale=0.75]{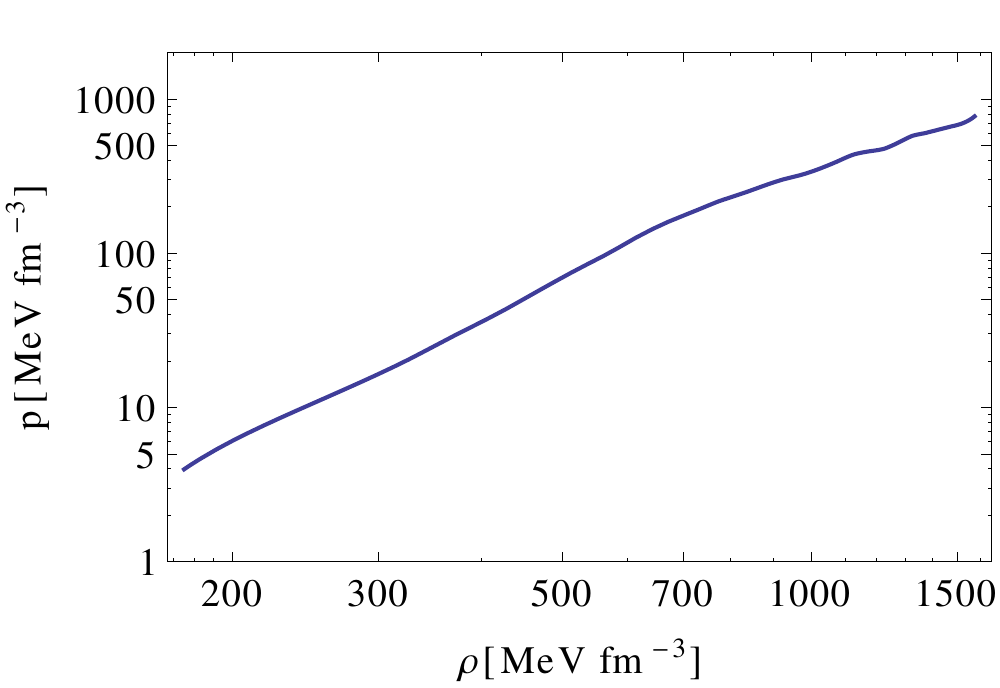}  &
\includegraphics[scale=0.55]{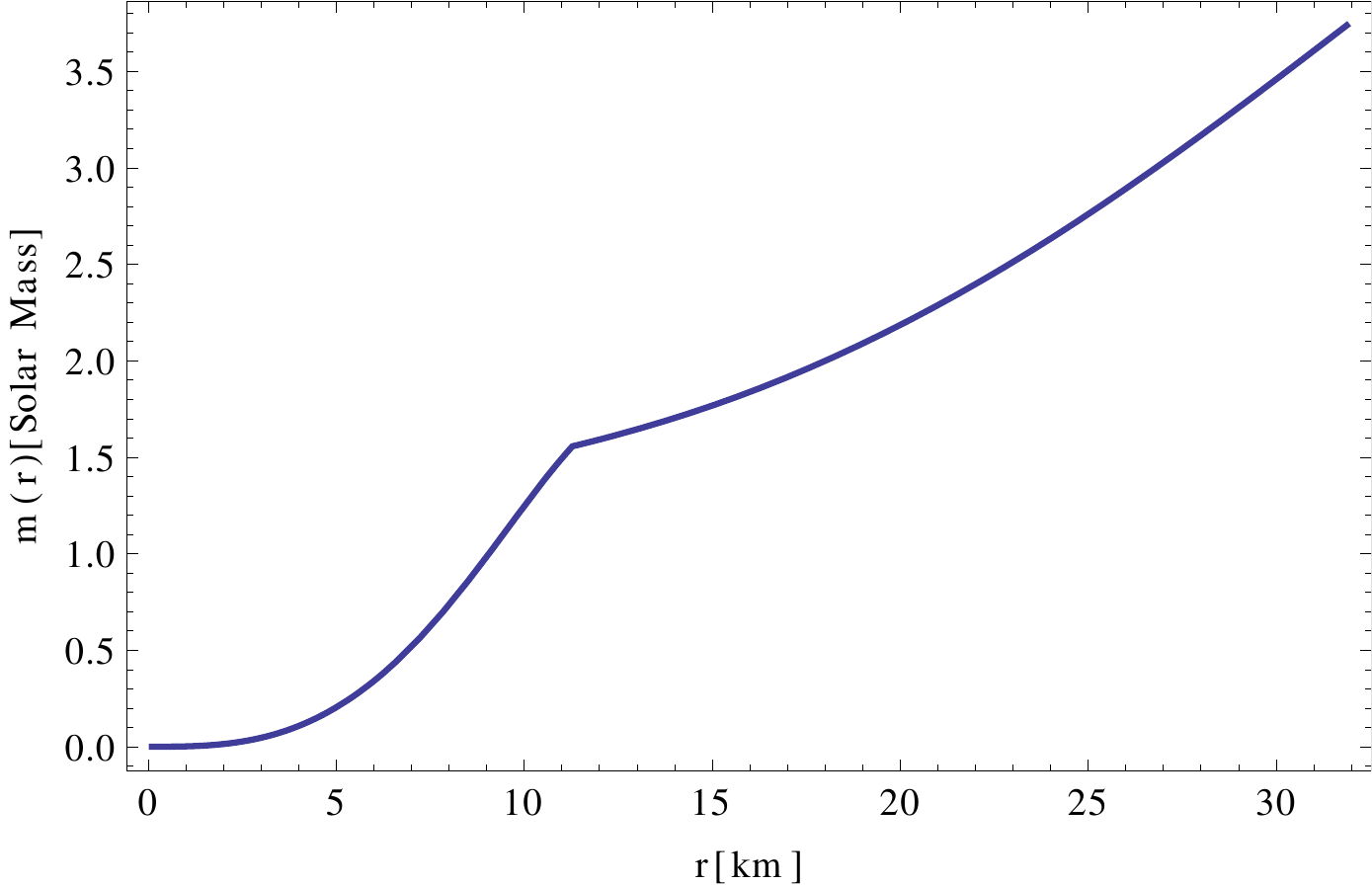} 
\end{array}$
\end{center} 
\caption{Left: pressure as function of energy density. Units for both are $MeV\ fm^{-3}$. Right:Number density times mass in  units of $g\ cm^{-3}$ as function of energy density in $MeV\ fm^{-3}$ } 
 \label{Fig:pn_rho}
\end{figure}
It was obtained by fitting observational data of x-ray bursters to study the mixed neutron star. The dependence of the pressure and the number density on the energy density are displayed  in Fig. \ref{Fig:pn_rho}.  
  Thus, the maximal  ordinary neutron star mass for this EOS is found to be $2.44 M_{\odot}$ and the  corresponding radius is $11.7 \ km$.

Let us assume that $\Lambda' =\frac{1}{2}\Lambda$ where  $\Lambda'$ and   $\Lambda$ are the scales for the mirror and ordinary  QCD,  respectively.  Since these scales largely control all masses, we expect that also  the mass of the dark fermion, $m_{b2}$ is half of the mass of the ordinary neutron, $m_{b1}$. 
For the dark baryons we use the same EOS scaled appropriately, so that the energy densities and pressures scale as the fourth power of the corresponding masses. Thus, with  this EOS, we have 
 \begin{equation}
 \label{p2rho2}
   p_2(\rho_2)=\frac{1}{16}  p_1 (16 \rho_2 )  
 \end{equation}
  
The  maximal mass of a  pure dark neutron star  is  $\sim 10 M_{\odot}$ and the corresponding radius  is     $\sim 50 km$.  It is expected that  the mixed neutron star solution  would yield    a   mass, and radius  intermediate   between those for a neutron star and a pure dark neutron star.  
  We checked that the numerical results indeed obey equation (\ref{constr}). We also found that in accord with equation (\ref{constr}),  indeed no static solutions are obtained when  $\rho_1(0) > 16 \rho_2(0)$.   
  
   We present an illustrative example of a typical mixed neutron star model for which:
$ \rho_1 (0)= 600\ {\rm MeV\ fm}^{-3}  \ ,\  {\rm and} \ \  \rho_2 (0)= \frac{1300}{16}    \  {\rm MeV\ fm}^{-3} $. 
 
\begin{figure}[ht]
\begin{center} $
\begin{array}{cc}
\includegraphics[scale=0.55 ]{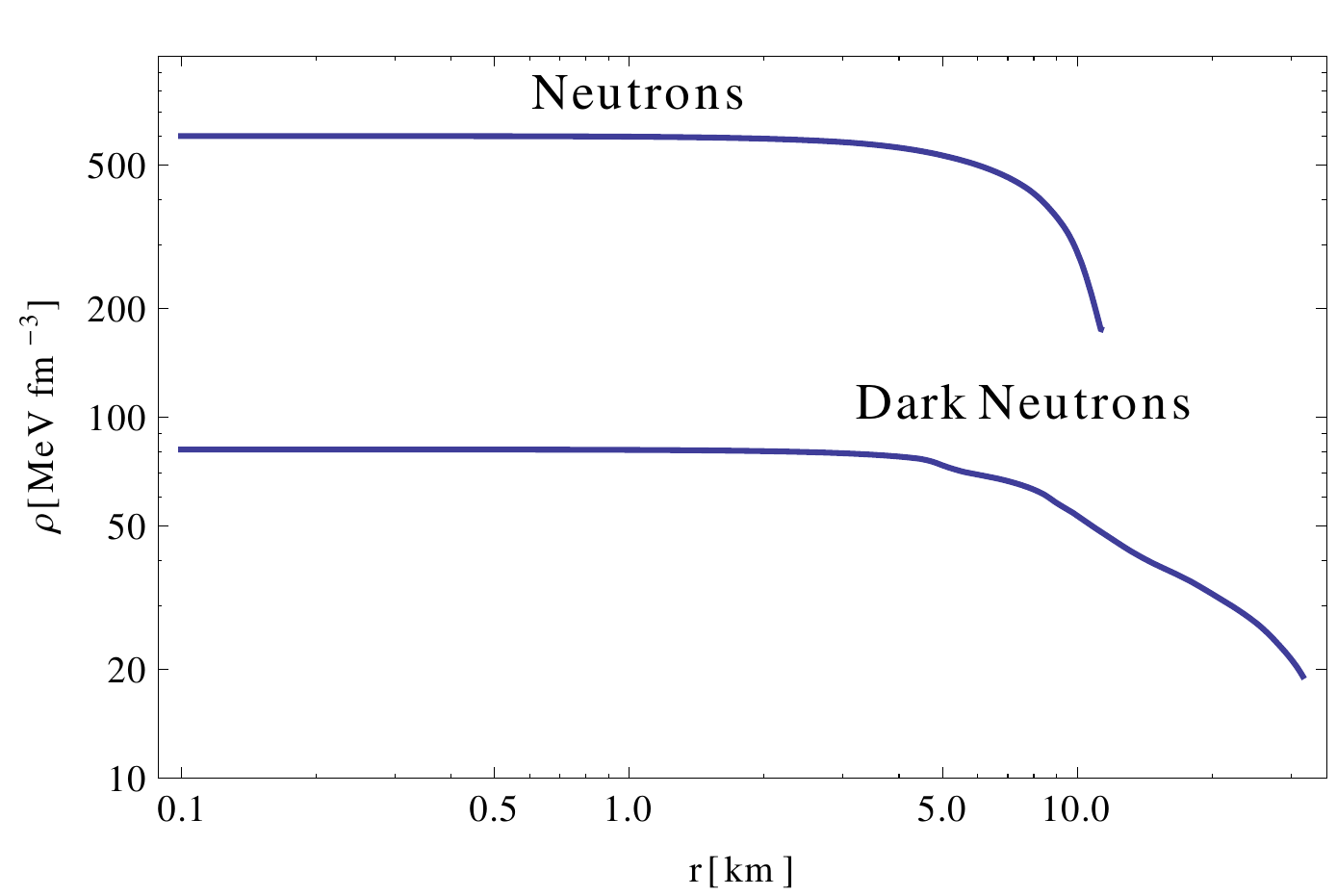}   &
 \includegraphics[scale=0.55]{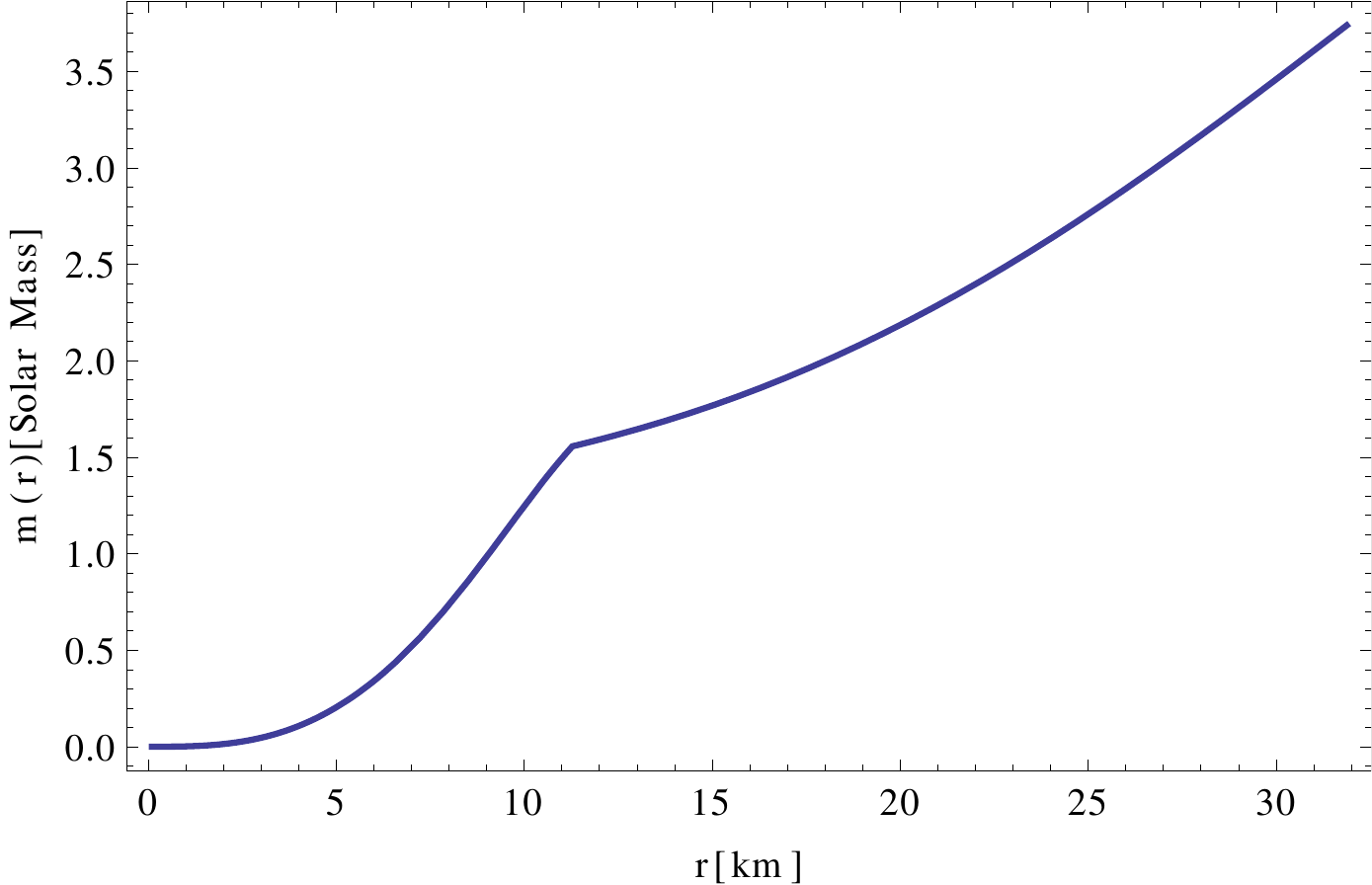}
\end{array}$
\end{center} 
\caption{Left:  energy densities as function of radial distance.   Right: Enclosed mass   as function of the radial distance. } 
 \label{Fig:rho12m}
 \end{figure}

\begin{figure}[h] 
\centerline{\includegraphics[scale=0.6]{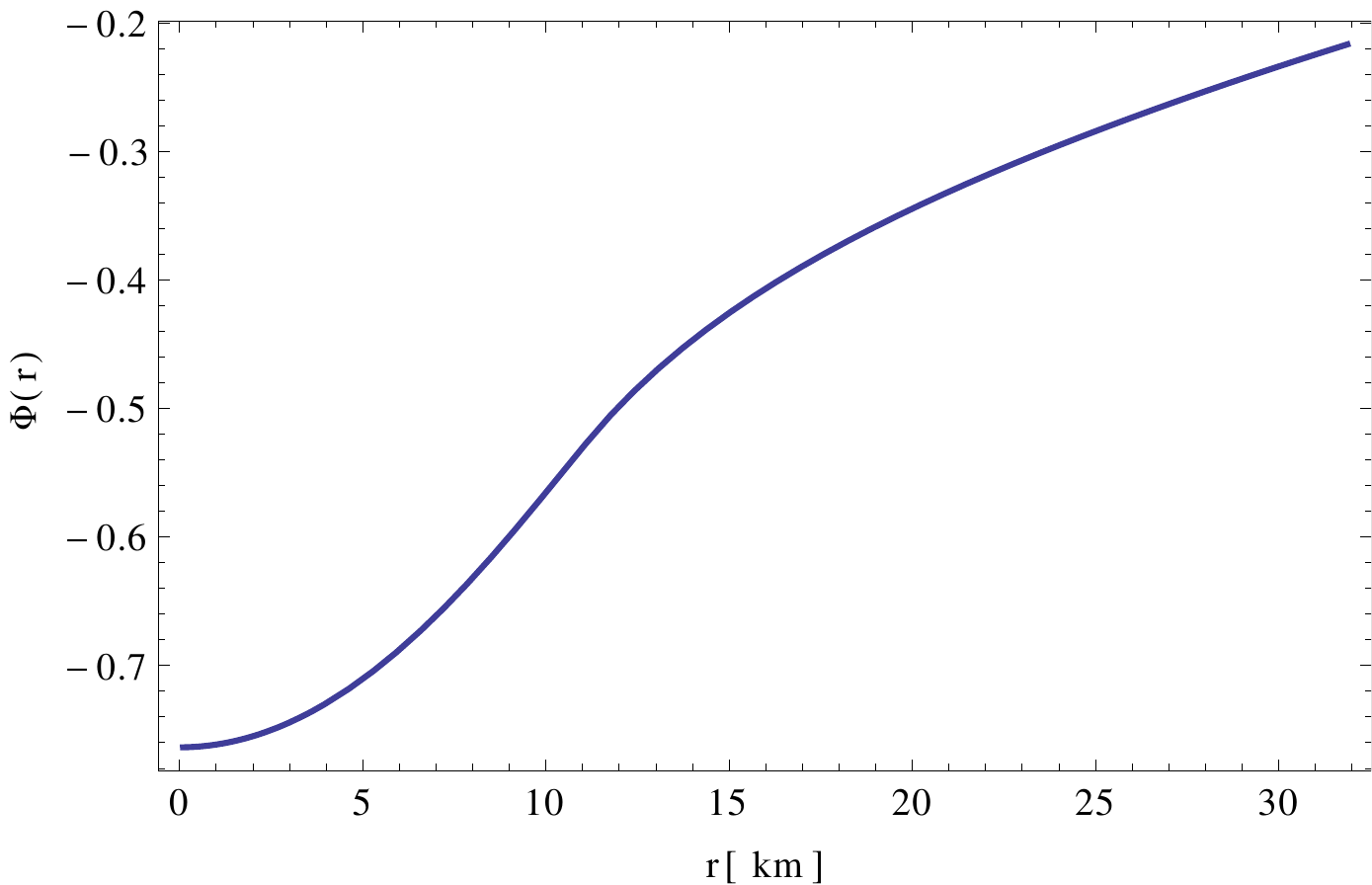}}
 \caption{$  \phi(r) $ as function of radius. 
  A photon emitted from the neutron surface $R_1$, is  redshifted by a factor 1.72 and a photon emitted from $R_2$ by 1.25.}
   \label{phi_r}
\end{figure}

 The results of the  computation are summarized in Table \ref{tbl}. The r-dependence of the energy densities, the enclosed mass $m(r)$, and $\phi(r)$ are displayed in figures (\ref{Fig:rho12m}),  (\ref{phi_r}).

\vskip -0.2  truecm
 \begin{table}[h]
%\begin{minipage}[]{140mm}
\caption{Model Results. The last entry is the gravitational binding energy of the neutrons  inside $R_1$,
divided by the total neutron mass  inside $R_1$: $(N_1 m_b - M_1)/((N_1 m_b)$}
\label{tbl}
     %\vskip 0.5 truecm 
\begin{center} 
\begin{tabular} { || p{1.6 cm}     p{1.6  cm} p{5.5 cm}     p{3.2 cm}||   } 
 \hline  \hline
M & $M_2$  &\ \ \ \ \ \ \ \ \ \  $M_1$ & $m(R_1)$
 \\   \hline
   $3.74 M_{\odot}$& $2.4 M_{\odot}$  &\ \ \ \ \ \ \ \ \ \ $1.34 M_{\odot}$ &   $1.56 M_{\odot}$\\
    \\ \hline  \hline
 $R_2$&
  $R_1$  & \ \ \ \ \ \ \  Redshifts & Neutron BE  \\
  \hline   
      31.9    km &  11.1 km    & \ \ \ \  $z(R_1)$=0.72,\   $z(R_2)$=0.25    & \ \ \ \ 22\% \\ \hline
    \end{tabular} 
   \end{center}

 \end{table}

% \newpage
\section{Discussion and comments}
The above example  demonstrates that a mixed neutron star can, as expected, have
a total mass higher as measured by, say, orbital dynamics than pure ordinary neutron stars. At the same time the radius, as probed by ordinary massless and massive particles, is the neutron-sphere radius which is similar in value to the radius of ordinary neutron stars as is the mass as measured by red shift analysis.

There are  interesting  implications for phenomenology of compact X-ray sources, related to the modified  redshifts  of  emitted photons. % from the neutron-sphere is about  50\% higher than in the ordinary neutron star case. 
 Also, since the maximal neutron mass stable against collapse to a black hole  would be smaller, than in the pure neutron case,  an accretion induced collapse to a black hole would be more likely to occur. %This will be followed by accretion of the outer envelope of the dark neutrons.
 
 The main objective of this paper is to demonstrate that  mixed neutron star with masses  exceeding those of ordinary neutron  star  are possible and to study some of their general features.  We now comment briefly  on the scenarios that can lead to the formation of such compact objects.

% \subsection*{Possible origin of composite "mixed neutron star" in ADM models}

%Roughly equally mixed   neutron  stars    can form only in regions where ordinary and dark matter have jointly clustered . 
Asymmetric dark matter can cluster without self-annihilation. 
Still in order for joint clustering to actually happen, further specific features are needed which may be  difficult to build into complete consistent models.
We will not discuss in detail how mixed stars may evolve but only sketch in broad
terms how roughly equal masses of order of solar mass of ordinary and dark baryons may possibly be brought together.
Two general  scenarios can be envisioned: 
\begin{enumerate}

\item  Dark matter is accreted onto ordinary stellar objects at various stages of the evolution of the latter, or conversely,  ordinary matter is accreted onto pre-existing dark stars , and 

\item  Dark and ordinary matter  jointly cluster  forming the mixed stars
\end{enumerate}

\noindent 1.  Accretion of DM onto stars has been discussed in the past \cite{spergel,Gilliland,sarkar}. It was motivated by noting that even tiny
 ( $ \eta(X) =N_X/N_{Baryons}$  $\sim 10^{-11}$ ) admixtures inside the sun of dark matter of mass  $m(X) \sim 5-10$ GeV  can modify heat convection from the 
 solar core and help explain some apparent anomalies.
If the density of dark matter near the star has the average value of $  0.4~ Gev/{m_X}$ , then to generate $\eta(X) \sim10^{-11}$  over a Hubble time we need that $\sigma(X-N)$, the cross- section for scattering of dark and normal nucleons, exceed $10^{-37} cm^2$. This is  excluded for heavy DM by direct searches - but not for the case of m(X) =1/2 GeV that we focus on in this paper. Furthermore once $\eta(X)$  exceeds the ratio of $\sigma_{XN}/\sigma_{XX}$ (which   can be as small as $10 ^{-15}$), the nonlinear process of scattering the incoming DM  on already captured X particles in the star dominates and  further accelerates the accretion~\cite{sarkar}. There is however an upper ``unitarity"  limit on the accretion rate fixed by the area of the star $\pi R^2$  (  possibly with a ``focusing" $(v_{escape} /{v_{virial} })^2$ enhancement $\sim$10 for the sun ) corresponding to the case of complete capture of all X particles which hit the  stellar surface.  

  Even at this maximal rate  if a solar type star were to accrete in Hubble time a solar mass of dark matter, we need that the DM density in its neighborhood will be $10^9$  times larger than the local halo density of 0.4 GeV cm$^{-3}$.  In general CDM starts clustering before baryons and our star may naturally be situated in a dark matter mini-halo. If this dark mini-halo formed at redshift z its density can be enhanced in comparison with the cosmological DM density  of ~ $ KeV/{cm^3}$ by $(6z) ^3 \sim 2\times 10^8$ for $z\sim100$. The CMB spectrum and simulations~\cite{moore}, certainly exclude forming mini-halos of solar mass at larger redshifts.  Even then this achieves at most a $10^3$ enhancement relative to the local halo density. It seems that only if DM was dissipative it could have clustered more effectively reaching the $10^9$ enhancement required.

Mirror DM can be dissipative  if the mirror photon mass $m_{\gamma'}$  is smaller than the atomic mirror excitations $O(m_{e^\prime} \alpha^2)$ where $e^\prime$ is the mirror electron. Yet we should  keep $m_{\gamma'} > 2m_e$. Otherwise $\gamma'$ becomes stable and can escape the collider leaving a missing energy signature in CMS  or many other fixed target experiments where dark photons have been searched \cite{Bjorken}. %This  could be avoided if  the coupling to ordinary electrons and quarks etc is strongly suppressed - which in turn  would limit the X-N cross sections.

 In passing, we note that the total DM accretion  is not enhanced for bigger, more massive red/blue giant stars . The surface density~ $M/{R^2}$ of the such giant stars is smaller making it difficult to accrete the minimal amount of X particles required in order to initiate the non-linear regime and hence reach the unitarity limit . Furthermore the lifetime of these stars scales like  $M^{-3}$ making them live considerably shorter than solar systems. Also for more compact objects such as white dwarfs /neutron stars the enhanced focusing % F (~ 1/
  is offset by the far smaller  areas.       

\noindent 2. This leaves us with the second scenario where dark matter, while strongly interacting on its own,  is non-dissipative . The idea is that the dissipative baryons that fall
 into the initial potential wells generated by the DM, will through gravitational interactions with the DM, dissipate also its energy. This then can allow dark and ordinary matter to jointly co-cluster into denser and denser structures so as to form eventually the mixed stars .

\section{Summary}
In summary, we have investigated the question of maximum neutron star mass if a substantial fraction of its mass is contributed by fermionic dark matter particles. We do this by solving the relativistic TOV equations using similar equations of state for ordinary baryonic and dark baryonic matter as well as with heuristic consideration of balancing kinetic thermal energy with gravitational energy of the two components (dark and ordinary baryons) of the neutron star.  We find examples where for the dark matter mass being half the neutron mass, leads to neutron star mass two to four times higher than Chandrasekhar mass. We also comment on possible scenarios where the required initial conditions for the abundances of the dark matter in the neutron star could arise. This work should be considered as a beginning attempt to get some ideas about the complex problem of two strongly interacting dark fermion species in a compact star and is meant to inspire future works on the subject. After a summary of our work appeared~\cite{Goldman_11}, this problem was also considered in \cite{li} where it was argued that the presence of dark matter inside a neutron star softens the equation of state more strongly than hyperons, thereby changing its mass.

\section{Acknowledgement} The work of R. N. M. is supported by National Science Foundation grant No. PHY-0968854.

\end{document}